\documentclass[global]{svjour}
\usepackage{amssymb}
\usepackage{amsfonts}
\usepackage{amsmath}

\input{tcilatex}

\begin{document}

\title{Parabolic Equations and Markov Processes Over $p-$adic Fields}
\author{W. A. Z\'{u}\~{n}iga-Galindo \thanks{%
Project sponsored by the National Security Agency under Grant Number
H98230-06-1-0040. The United States Government is authorized to reproduce
and distribute reprints notwithstanding any copyright notation herein. \ \ \
\ \ \ \ \ \ \ \ \ \ \ \ \ \ \ \ \ \ \ \ \ \ \ \ \ \ AMS Classification.
Primary 35K30; Secondary 46S10.}}
\institute{Centro de Investigaci\'{o}n y de Estudios Avanzados del I.P.N.\\
Departamento de Matem\'{a}ticas\\
Av. Instituto Polit\'{e}cnico Nacional 2508\\
Col. San Pedro Zacatenco, M\'{e}xico D.F., C.P. 07360\\
M\'{e}xico\\
Barry University, Department of Mathematics and Computer Science\\
11300 NE Second Avenue, Miami Shores, FL 33161, U.S.A.\\
%TCIMACRO{\TeXButton{email}{\email{wzuniga@math.cinvestav.mx}} }%
%BeginExpansion
\email{wzuniga@math.cinvestav.mx}
%EndExpansion
}
\maketitle

\begin{abstract}
In this paper we construct and study a fundamental solution of Cauchy's
problem for $p-$adic parabolic equations of the type 
\begin{equation*}
\frac{\partial u\left( x,t\right) }{\partial t}+\left( f\left( D,\beta
\right) u\right) \left( x,t\right) =0,x\in \mathbb{Q}_{p}^{n},n\geq 1,t\in
\left( 0,T\right] ,
\end{equation*}%
where $f\left( D,\beta \right) $, $\beta >0$, is an elliptic
pseudo-differential operator. We also show that the fundamental solution is
the transition density of a Markov process with state space $\mathbb{Q}%
_{p}^{n}$.
\end{abstract}

\keywords{Parabolic equations, ultrametric diffusion, pseudo-differential
operators, Markov processes, $p-$adic fields.}

\section{Introduction}

In recent years $p-$adic analysis has received a lot of attention due to its
applications in mathematical physics, see e.g. \cite{A-K1}, \cite{A-K2}, 
\cite{A-B-K-O}, \cite{A-B-O}, \cite{KH1}, \cite{KH2}, \cite{KO2}, \cite{R-T}%
, \cite{V-V-Z}\ and references therein. All these developments have been
motivated for two physical ideas. The first \ is the conjecture in particle
physics that at Planck distances the space-time has a non-Archimedean
structure. As a consequence of this idea have emerged the $p-$adic quantum
mechanics and $p-$adic quantum field theory, see e.g. \cite{KH1}, \cite{Va}, 
\cite{V-V-Z}. The second idea comes from statistical \ physics, \ in
particular in connection with models describing relaxation in glasses,
macromolecules, and proteins. It has been proposed that the non exponential
nature of those relaxations is a consequence of a hierarchical structure of
the state space which \ can in turn \ be put in connection with $p-$adic
structures, see e.g. \cite{A-B-K-O}, \cite{A-B-O}, \cite{R-T}.

In this paper, motivated by \cite{A-B-K-O} and \cite{KO1}, we consider the
Cauchy\ problem%
\begin{equation}
\left\{ 
\begin{array}{ll}
\frac{\partial u\left( x,t\right) }{\partial t}+\left( f\left( D,\beta
\right) u\right) \left( x,t\right) =0\text{,} & x\in \mathbb{Q}_{p}^{n}\text{%
, }n\geq 1\text{, }t\in \left( 0,T\right] \\ 
&  \\ 
u\left( x,0\right) =\varphi \left( x\right) , & 
\end{array}%
\right.  \label{Cauchy}
\end{equation}%
where $\mathbb{Q}_{p}$ is the field of $p-$adic numbers, $f\left( D,\beta
\right) $ is an elliptic pseudo-differential operator of the form 
\begin{equation*}
\left( f\left( D,\beta \right) \phi \right) \left( x,t\right) =\mathcal{F}%
_{\xi \rightarrow x}^{-1}\left( \left\vert f\left( \xi \right) \right\vert
_{p}^{\beta }\mathcal{F}_{x\rightarrow \xi }\phi \left( x,t\right) \right) .
\end{equation*}%
Here $\mathcal{F}$ denotes the Fourier transform, $\beta $ is a positive
real number, and $f\left( \xi \right) \in \mathbb{Q}_{p}\left[ \xi
_{1},\ldots ,\xi _{n}\right] $ is a homogeneous polynomial of degree \ $d$
satisfying the property 
\begin{equation*}
f\left( \xi \right) =0\text{ if and only if }\xi =0.
\end{equation*}%
We establish the existence of a unique solution to Problem \ref{Cauchy} in
the case in which $\varphi \left( x\right) $ is a continuous and an
integrable function. Under these hypotheses we show the existence of a
solution $u\left( x,t\right) $ that is continuous in $x$, for a fixed $t\in %
\left[ 0,T\right] $, bounded, and integrable function (see Theorem \ref%
{prop3}). In addition the solution can be presented in the form

\begin{equation*}
u\left( x,t\right) =Z\left( x,t\right) \ast \varphi \left( x\right)
\end{equation*}%
where $Z\left( x,t\right) $ is\textit{\ the fundamental solution }(also
called \textit{the heat kernel}) to Cauchy's Problem \ref{Cauchy}:%
\begin{equation}
Z\left( x,t,f,\beta \right) :=Z\left( x,t\right) =\tint\limits_{\mathbb{Q}%
_{p}^{n}}\Psi \left( x\cdot \xi \right) e^{-t\left\vert f\left( \xi \right)
\right\vert _{p}^{\beta }}d\xi .  \label{fusol}
\end{equation}%
Here $\Psi \left( \cdot \right) $ is \ a standard additive character on $%
\mathbb{Q}_{p}$, $x\cdot \xi :=\tsum\nolimits_{i=1}^{n}x_{i}\xi _{i}$, for $%
x $, $\xi \in \mathbb{Q}_{p}^{n}$, and $t\in \left( 0,T\right] $. The
fundamental solution is a transition density of a Markov process \ with
space state $\mathbb{Q}_{p}^{n}$ (see Theorem \ref{theo3}). The study of the
decay of the oscillatory integral $Z\left( x,t\right) $ as $\left\Vert
x\right\Vert _{p}:=\max_{i}\left\vert x_{i}\right\vert _{p}$ approaches
infinity, for any fixed $t>0$, plays a central role in this article (see
Theorem \ref{theo1}). It is relevant to mention that Igusa developed \ a
method for estimating a large class of $p-$adic\ oscillatory integrals \cite%
{Ig1}, \cite{Ig2}, but this method is not applicable to $Z\left( x,t\right) $%
. Our analysis of the decay of the fundamental solution uses some geometric
ideas developed \ by the author to study local zeta functions of
semiquasi-homogeneous polynomials \cite{Z-G}. The fundamental solution is a
non-negative function, i.e., $Z\left( x,t\right) \geq 0$, for every $x\in 
\mathbb{Q}_{p}^{n}$, $t>0$, (see Theorem \ref{theo1A}). In the proof of this
crucial property we use integration over fibers which is a technique \
typically used in the study of \ local zeta functions (see e.g. \cite[Sect.
7.6]{Ig2}).

In dimension one the $p-$adic heat kernel has been studied extensively \cite%
{Bla}, \cite{Ha1}, \cite{Ha2}, \cite{I}, \cite{KO1}, \cite{KO2}, \cite{V-V-Z}%
, A systematic study of the parabolic equations and the related Markov
processes was given by Kochubei in \cite{KO1}-\cite{KO2}. In \cite{Va}
Varadarajan studied the heat kernel on $n-$dimensional vector spaces over
local fields and division rings. Our results constitute a first step toward
the generalization of the results of \cite{KO1} and \cite{Va} to higher
dimension.

The author thanks to the referee for his or her useful comments, which led
to an improvement of this work.

\section{\label{Prem}Preliminaries}

We denote by $\mathbb{Q}_{p}$ the field of $p-$adic numbers and by $\mathbb{Z%
}_{p}$ the ring of $p-$adic integers. For $x\in \mathbb{Q}_{p}$, $ord\left(
x\right) \in \mathbb{Z}\cup \left\{ +\infty \right\} $ denotes the valuation
of $x$, and $\left\vert x\right\vert _{p}=p^{-ord\left( x\right) \text{ }}$%
its absolute value. We extend this absolute value to $\mathbb{Q}_{p}^{n}$ by
taking $\left\Vert x\right\Vert _{p}=\max_{i}\left\vert x_{i}\right\vert
_{p} $\ for $x=\left( x_{1},\ldots ,x_{n}\right) \in \mathbb{Q}_{p}^{n}$.

Let $\Psi :\mathbb{Q}_{p}\rightarrow \mathbb{C}^{\times }$ be the additive
character defined by%
\begin{equation*}
\begin{array}{lll}
\Psi :\mathbb{Q}_{p}\rightarrow \mathbb{Q}_{p}/\mathbb{Z}_{p}\hookrightarrow
& \mathbb{Q}/\mathbb{Z} & \rightarrow \mathbb{C}^{\times } \\ 
& b & \rightarrow \exp \left( 2\pi ib\right) .%
\end{array}%
\end{equation*}%
Let $V=\mathbb{Q}_{p}^{n}$ be the $n-$dimensional $\mathbb{Q}_{p}-$vector
space and $V^{\prime }$ its algebraic dual vector space. We identify $%
V^{\prime }$ with $\mathbb{Q}_{p}^{n}$ via the $\mathbb{Q}_{p}-$bilinear form%
\begin{equation*}
x\cdot y=x_{1}y_{1}+\ldots +x_{n}y_{n},
\end{equation*}%
$x\in V=\mathbb{Q}_{p}^{n}$, $y\in V^{\prime }=\mathbb{Q}_{p}^{n}$. Now we
identify $V^{\prime }$ with the topological dual $V^{\ast }$ of $V$ (i.e.
the group of all continuous additive characters on $\left( V,+\right) $) as
right $\mathbb{Q}_{p}-$vector spaces by means of the pairing $\Psi \left(
x\cdot y\right) $. The Haar measure $dx$ \ is autodual with respect this
pairing.

We denote \ the space of all Schwartz-Bruhat functions on $\mathbb{Q}%
_{p}^{n} $ by $\mathcal{S}=\mathcal{S}\left( \mathbb{Q}_{p}^{n}\right) $.
For $\phi \in \mathcal{S}$, we define its Fourier transform $\mathcal{F}\phi 
$ by 
\begin{equation*}
\left( \mathcal{F}\phi \right) \left( \xi \right) =\tint\limits_{\mathbb{Q}%
_{p}^{n}}\Psi \left( -x\cdot \xi \right) \phi \left( x\right) dx.
\end{equation*}%
Then the Fourier transform induces a linear isomorphism of $\mathcal{S}$\
onto $\mathcal{S}$ and the inverse Fourier transform is given by%
\begin{equation*}
\left( \mathcal{F}^{-1}\varphi \right) \left( x\right) =\tint\limits_{%
\mathbb{Q}_{p}^{n}}\Psi \left( x\cdot \xi \right) \varphi \left( \xi \right)
d\xi ,
\end{equation*}%
for $\varphi \in \mathcal{S}$. The map $\phi \rightarrow \mathcal{F}\phi $
is an $L^{2}$-isometry on $L^{1}\mathbb{\cap }L^{2}$, which is a dense
subspace of $L^{2}$.

\subsection{Elliptic Pseudo-differential Operators}

Let $h\left( \xi \right) \in \mathbb{Q}_{p}\left[ \xi _{1},\ldots ,\xi _{n}%
\right] $ be a non-constant polynomial. An operator of the form $h\left(
D,\beta \right) \phi =\mathcal{F}^{-1}\left( \left\vert h\right\vert
_{p}^{\beta }\mathcal{F}\phi \right) ,$ $\beta >0$, $\phi \in \mathcal{S}$,
is called a pseudo-differential operator with symbol $\left\vert
h\right\vert _{p}^{\beta }=\left\vert h\left( \xi \right) \right\vert
_{p}^{\beta }$. The operator $h\left( D,\beta \right) $ has a self-adjoint
extension with dense domain in $L^{2}$.

\begin{definition}
Let $f\left( \xi \right) \in \mathbb{Q}_{p}\left[ \xi _{1},\ldots ,\xi _{n}%
\right] $ be a non-constant polynomial. We say that $f\left( \xi \right) $
is an elliptic polynomial of degree $d$, if it satisfies: (1) $f\left( \xi
\right) $ is a homogeneous polynomial of degree $d$, and (2) $f\left( \xi
\right) =0$ $\Leftrightarrow $ $\xi =0$.
\end{definition}

\begin{lemma}
\label{lem1}Let $f\left( \xi \right) \in \mathbb{Q}_{p}\left[ \xi \right] $, 
$\xi =\left( \xi _{1},\ldots ,\xi _{n}\right) $, be an elliptic polynomial
of degree $d$. Then there exist positive constants $C_{0}=C_{0}(f)$, $%
C_{1}=C_{1}(f)$ such that 
\begin{equation*}
C_{0}\left\Vert \xi \right\Vert _{p}^{d}\leq \left\vert f\left( \xi \right)
\right\vert _{p}\leq C_{1}\left\Vert \xi \right\Vert _{p}^{d}\text{, for
every }\xi \in \mathbb{Q}_{p}^{n}.
\end{equation*}
\end{lemma}

\begin{proof}
Without loss of generality we may assume that $\xi \neq 0$. Let $\widetilde{%
\xi }\in \mathbb{Q}_{p}^{\times }$ be an element such that $\left\vert 
\widetilde{\xi }\right\vert _{p}=\left\Vert \xi \right\Vert _{p}\neq 0$. We
first note that%
\begin{equation}
\left\vert f\left( \xi \right) \right\vert _{p}=\left\vert \widetilde{\xi }%
\right\vert _{p}^{d}\left\vert f\left( \widetilde{\xi }^{-1}\xi \right)
\right\vert _{p}\text{,}  \label{iden1}
\end{equation}%
with $\widetilde{\xi }^{-1}\xi \in A:=\left\{ \left( z_{1},\ldots
,z_{n}\right) \in \mathbb{Z}_{p}^{n}\mid \left\vert z_{i}\right\vert _{p}=1%
\text{, for some }i\right\} $. Since $\left\vert f\right\vert _{p}$ is
continuous on $A$, that is a compact subset of $\mathbb{Z}_{p}^{n}$, it
verifies that $\inf_{z\in A}\left\vert f\left( z\right) \right\vert _{p}$,
and $\sup_{z\in A}\left\vert f\left( z\right) \right\vert _{p}$ are attained
on $A$, and since $\left\vert f\right\vert _{p}>0$ on $A$, we have 
\begin{equation*}
\sup_{z\in A}\left\vert f\left( z\right) \right\vert _{p}\geq \inf_{z\in
A}\left\vert f\left( z\right) \right\vert _{p}=q^{-m\left( f,A\right) }\text{%
, with }m\left( f,A\right) \in \mathbb{Z}\text{.}
\end{equation*}%
Therefore $0<\inf_{z\in A}\left\vert f\left( z\right) \right\vert _{p}\leq
\left\vert f\left( \widetilde{\xi }^{-1}\xi \right) \right\vert _{p}\leq
\sup_{z\in A}\left\vert f\left( z\right) \right\vert _{p}$, for $\widetilde{%
\xi }^{-1}\xi $ $\in A$. Now the result follows from (\ref{iden1}), 
\begin{equation*}
\left( \inf_{z\in A}\left\vert f\left( z\right) \right\vert _{p}\right)
\left\vert \widetilde{\xi }\right\vert _{p}^{d}\leq \left\vert f\left( \xi
\right) \right\vert _{p}\leq \left( \sup_{z\in A}\left\vert f\left( z\right)
\right\vert _{p}\right) \left\vert \widetilde{\xi }\right\vert _{p}^{d}.
\end{equation*}%
$\square $
\end{proof}

From now on $f\left( \xi \right) $ will denote an elliptic polynomial of
degree $d$. Since $cf\left( \xi \right) $ is elliptic for any $c\in \mathbb{Q%
}_{p}^{\times }$ when $f\left( \xi \right) $ is elliptic, we will assume
that all the elliptic polynomials have coefficients in $\mathbb{Z}_{p}$.

\begin{definition}
If $f\left( \xi \right) \in \mathbb{Z}_{p}\left[ \xi \right] $ is an
elliptic polynomial of degree $d$, then we say that $\left\vert f\right\vert
_{p}^{\beta }$\ is an elliptic symbol, and that $f\left( D,\beta \right) $
is an elliptic pseudo-differential operator of order $d$.
\end{definition}

\begin{remark}
Note that $g\left( \xi _{1},\xi _{2}\right) =\xi _{1}^{d}+p\xi _{2}^{d}$, $%
d\geq 2$, is an elliptic polynomial of degree $d$. In addition, if $l\left(
\xi \right) ,h\left( \xi \right) \in \mathbb{Z}_{p}\left[ \xi \right] $ are
elliptic polynomials then $F\left( \xi \right) =\left[ l\left( \xi \right) %
\right] ^{2}+p\left[ h\left( \xi \right) \right] ^{2}$ satisfies $F\left(
\xi \right) =0\Leftrightarrow \xi =0$. Therefore, given $m\geq 1$, there
exists an elliptic polynomial in $m$ variables. The elliptic quadratic forms
have been extensively studied, see e.g. \cite[Chapter 1]{B-S}. It is known
that all quadratic forms in five or more variables are not elliptic.
\end{remark}

\section{Basic Properties of the Fundamental Solution}

\subsection{Decaying of the Fundamental Solution at Infinity}

\begin{lemma}
\label{lem2}For every $t>0$, $\left\vert Z\left( x,t\right) \right\vert \leq
Ct^{\frac{-n}{d\beta }}$, where $C$ is a positive constant. Furthermore, $%
\Psi \left( x\cdot \xi \right) e^{-t\left\vert f\left( \xi \right)
\right\vert _{p}^{\beta }}\in L^{1}$ as a function of $\xi $, for every $%
x\in \mathbb{Q}_{p}^{n}$, and $t>0$.
\end{lemma}

\begin{proof}
Let an integer $m$ be such that $p^{m-1}\leq \left( C_{0}t\right) ^{\frac{1}{%
d\beta }}\leq p^{m}$. By applying Lemma \ref{lem1} we have%
\begin{eqnarray*}
\left\vert Z\left( x,t\right) \right\vert &\leq &\tint\limits_{\mathbb{Q}%
_{p}^{n}}e^{-C_{0}t\left\Vert \xi \right\Vert _{p}^{d\beta }}d\xi \leq
\tint\limits_{\mathbb{Q}_{p}^{n}}e^{-p^{\left( m-1\right) d\beta }\left\Vert
\xi \right\Vert _{p}^{d\beta }}d\xi \\
&\leq &p^{n}C_{0}^{-\frac{n}{d\beta }}\left( \tint\limits_{\mathbb{Q}%
_{p}^{n}}e^{-\left\Vert z\right\Vert _{p}^{d\beta }}dz\right) t^{-\frac{n}{%
d\beta }},
\end{eqnarray*}%
where $C_{0}$ is a positive constant. Since $e^{-\left\Vert z\right\Vert
_{p}^{d\beta }}\leq \left\Vert z\right\Vert _{p}^{-d\beta M}$, for any $M\in 
\mathbb{N}\setminus \left\{ 0\right\} $, and $\left\Vert z\right\Vert _{p}$
big enough, \ we can choose an $M$ such that $d\beta M=n+\epsilon $, $%
\epsilon >0$, and then 
\begin{equation*}
\tint\limits_{\left\Vert z\right\Vert _{p}\geq p^{r}}e^{-\left\Vert
z\right\Vert _{p}^{d\beta }}dz\leq \tint\limits_{\left\Vert z\right\Vert
_{p}\geq p^{r}}\frac{1}{\left\Vert z\right\Vert _{p}^{n+\epsilon }}dz<\infty
.
\end{equation*}%
Here $r$ is a positive constant depending on $M$. Therefore there exists a
positive constant $\ C$ such that%
\begin{equation*}
\left\vert Z\left( x,t\right) \right\vert \leq Ct^{-\frac{n}{d\beta }},
\end{equation*}%
for every $t>0$. $\square $
\end{proof}

\begin{lemma}
\label{lem3}Let $A$ be a compact subset of $\mathbb{Q}_{p}^{n}$ such that $%
0\notin A$. Then there exists a non-zero constant $M=M(f,A)\in \mathbb{N}$
such that $\left\vert f\left( \xi \right) \right\vert _{p}\geq p^{-M}$, for
every $\xi \in A$. In addition, for every ball $B=z+\left( p^{M+1}\mathbb{Z}%
_{p}\right) ^{n}\subset A$, it verifies that $\left\vert f\left( \xi \right)
\right\vert _{p}=\left\vert f\left( z\right) \right\vert _{p}$, for every $%
\xi \in B$.
\end{lemma}

\begin{proof}
In the proof of Lemma \ref{lem1} we showed the existence of an integer
constant $m=m(f,A)$ such that 
\begin{equation}
\left\vert f\left( \xi \right) \right\vert _{p}\geq p^{-m},\xi \in A.
\label{ineq}
\end{equation}%
If $m<0$, then $\left\vert f\left( \xi \right) \right\vert _{p}\geq
p^{-m}>p^{-1}$, for $\xi \in A$. Then without loss of generality we can
assume that\ (\ref{ineq}) holds for $m=m(f,A)\geq 1$.

We first consider the case $A\subseteq \mathbb{Z}_{p}^{n}$. We take $%
M=m(f,A) $. By applying the ultrametric triangle inequality and using (\ref%
{ineq}) we have 
\begin{equation*}
\left\vert f\left( z+p^{M+1}\eta \right) \right\vert _{p}=\left\vert f\left(
z\right) +p^{M+1}g_{z}\left( \eta \right) \right\vert _{p}=\left\vert
f\left( z\right) \right\vert _{p},
\end{equation*}%
for any $\eta \in \mathbb{Z}_{p}^{n}$, since $\left\vert g_{z}\left( \eta
\right) \right\vert _{p}\leq 1$.

Now we consider the case $A\nsubseteq $ $\mathbb{Z}_{p}^{n}$. For $\xi
=\left( \xi _{1},\ldots ,\xi _{n}\right) \in A$, we set $ord\left( \xi
\right) :=\min_{1\leq i\leq n}ord\left( \xi _{i}\right) $. Then $\xi =%
\widetilde{\xi }p^{ord\left( \xi \right) }$, with $\widetilde{\xi }\in 
\mathbb{Z}_{p}^{n}$, and $\left\Vert \xi \right\Vert _{p}=p^{-ord\left( \xi
\right) }$. By using that $A$ is compact and $A\nsubseteq $ $\mathbb{Z}%
_{p}^{n}$, we have $\sup_{\xi \in A}\left\Vert \xi \right\Vert
_{p}=p^{c\left( f,A\right) }$, with $c\left( f,A\right) $ a positive
integer, i.e., $ord\left( \xi \right) \geq -c\left( f,A\right) $, for every $%
\xi \in A$. Now applying (\ref{ineq}) we have 
\begin{equation}
\left\vert f\left( \widetilde{\xi }\right) \right\vert _{p}\geq
p^{-m+ord\left( \xi \right) d}\geq p^{-\left( m+c\left( f,A\right) d\right) }%
\text{, for every }\widetilde{\xi }\in p^{-c\left( f,A\right) }A\subseteq 
\mathbb{Z}_{p}^{n}.  \label{ineq-a}
\end{equation}%
Let $z=\widetilde{z}p^{ord(z)}\in A$. We assume that $ord(z)<0$, i.e., $%
z\notin \mathbb{Z}_{p}^{n}$. \ By taking $M=m+c\left( f,A\right) d$, and
applying the previous case to $p^{-c\left( f,A\right) }A$, that is a subset
of $\mathbb{Z}_{p}^{n}$, we have 
\begin{equation*}
\left\vert f\left( \widetilde{z}+p^{M+1}\eta \right) \right\vert
_{p}=\left\vert f\left( \widetilde{z}\right) +p^{M+1}g_{\widetilde{z}}\left(
\eta \right) \right\vert _{p}=\left\vert f\left( \widetilde{z}\right)
\right\vert _{p}\text{,}
\end{equation*}%
for $\widetilde{z}\in p^{-c\left( f,A\right) }A$, $\eta \in \mathbb{Z}%
_{p}^{n}$. Now by multiplying the above equality by $p^{-ord(z)d}$, 
\begin{equation*}
\left\vert f\left( z+p^{M+1+ord(z)}\eta \right) \right\vert _{p}=\left\vert
f\left( z\right) \right\vert _{p}\text{, for }\eta \in \mathbb{Z}_{p}^{n}.
\end{equation*}%
Finally, since $ord(z)<0$, $z+p^{M+1+ord(z)}\mathbb{Z}_{p}^{n}\supset
z+p^{M+1}\mathbb{Z}_{p}^{n}$, it follows from the previous equality that

\begin{equation*}
\left\vert f\left( z+p^{M+1}\eta \right) \right\vert _{p}=\left\vert f\left(
z\right) \right\vert _{p}\text{, for }\eta \in \mathbb{Z}_{p}^{n}.
\end{equation*}%
$\square $
\end{proof}

Define 
\begin{equation*}
Z_{L}\left( x,t,f,\beta \right) :=Z_{L}\left( x,t\right)
=\tint\limits_{\left( p^{-L}\mathbb{Z}_{p}\right) ^{n}}\Psi \left( x\cdot
\xi \right) e^{-t\left\vert f\left( \xi \right) \right\vert _{p}^{\beta
}}d\xi ,\text{ }L\in \mathbb{N},
\end{equation*}%
where $\beta >0$, $t>0$, and $f\left( \xi \right) \in \mathbb{Z}_{p}\left[
\xi _{1},\ldots ,\xi _{n}\right] $ is an elliptic polynomial of degree $d$.

\begin{lemma}
\label{lema4} If $\left\Vert x\right\Vert _{p}$ is big enough, and $%
p^{\left( M+1\right) d\beta }t\left\Vert x\right\Vert _{p}^{-d\beta }\leq 1$%
, then 
\begin{equation*}
\left\vert Z_{0}\left( x,t\right) \right\vert \leq Ct\left\Vert x\right\Vert
_{p}^{-d\beta -n},
\end{equation*}
where $C$ is a positive constant.
\end{lemma}

\begin{proof}
By applying Fubini's Theorem we have 
\begin{equation}
Z_{0}\left( x,t\right) =\tsum\limits_{l=0}^{\infty }\frac{\left( -1\right)
^{l}}{l!}t^{l}\tint\limits_{\mathbb{Z}_{p}^{n}}\Psi \left( x\cdot \xi
\right) \left\vert f\left( \xi \right) \right\vert _{p}^{\beta l}d\xi .
\label{I}
\end{equation}%
By using the fact that $\left\Vert x\right\Vert _{p}$ is big enough, we may
assume that $ord\left( x_{i_{0}}\right) <0$, for some $i_{0}$, then $%
\tint\nolimits_{\mathbb{Z}_{p}^{n}}\Psi \left( x\cdot \xi \right) d\xi =0$,
and (\ref{I}) can be rewritten as 
\begin{equation}
Z_{0}\left( x,t\right) =\tsum\limits_{l=1}^{\infty }\frac{\left( -1\right)
^{l}}{l!}t^{l}\tint\limits_{\mathbb{Z}_{p}^{n}}\Psi \left( x\cdot \xi
\right) \left\vert f\left( \xi \right) \right\vert _{p}^{\beta l}d\xi .
\label{I0}
\end{equation}%
We decompose $\mathbb{Z}_{p}^{n}$ as the disjoint union of $\left( p\mathbb{Z%
}_{p}\right) ^{n}$ and%
\begin{equation*}
A:=\left\{ \left( \xi _{1},\ldots ,\xi _{n}\right) \in \mathbb{Z}%
_{p}^{n}\mid ord\left( \xi _{i}\right) =0\text{, for some index }i\right\} ,
\end{equation*}%
and define 
\begin{equation*}
I\left( j,l\right) :=I\left( x,f,\beta ,j,l\right) =\tint\limits_{\mathbb{Z}%
_{p}^{n}}\Psi \left( p^{j}x\cdot \xi \right) \left\vert f\left( \xi \right)
\right\vert _{p}^{\beta l}d\xi ,\text{ for }j\geq 0\text{, }l\geq 1,
\end{equation*}%
and%
\begin{equation*}
\widetilde{I}\left( j,l,A\right) :=\widetilde{I}\left( x,f,\beta
,j,l,A\right) =\tint\limits_{A}\Psi \left( p^{j}x\cdot \xi \right)
\left\vert f\left( \xi \right) \right\vert _{p}^{\beta l}d\xi ,\text{ for }%
j\geq 0\text{, }l\geq 1.
\end{equation*}

Now we compute an expansion for $I\left( 0,l\right) $\ as follows:

\begin{eqnarray*}
I\left( 0,l\right) &=&\tint\limits_{\mathbb{Z}_{p}^{n}}\Psi \left( x\cdot
\xi \right) \left\vert f\left( \xi \right) \right\vert _{p}^{\beta l}d\xi \\
&=&\tint\limits_{\left( p\mathbb{Z}_{p}\right) ^{n}}\Psi \left( x\cdot \xi
\right) \left\vert f\left( \xi \right) \right\vert _{p}^{\beta l}d\xi
+\tint\limits_{A}\Psi \left( x\cdot \xi \right) \left\vert f\left( \xi
\right) \right\vert _{p}^{\beta l}d\xi \\
&=&p^{-n-\beta dl}I\left( 1,l\right) +\widetilde{I}\left( 0,l,A\right) .
\end{eqnarray*}%
By iterating this formula $k+1-$times we obtain that%
\begin{equation*}
I\left( 0,l\right) =\tsum\limits_{j=0}^{k}p^{-j\left( n+\beta dl\right) }%
\widetilde{I}\left( j,l,A\right) +p^{-\left( k+1\right) \left( n+\beta
dl\right) }I\left( k+1,l\right) .
\end{equation*}%
Hence $I\left( 0,l\right) $ admits the\ expansion 
\begin{equation}
I\left( 0,l\right) =\tsum\limits_{j=0}^{\infty }p^{-j\left( n+\beta
dl\right) }\widetilde{I}\left( j,l,A\right) ,  \label{I2a}
\end{equation}%
which is uniform in all the parameters. On the other hand, since $A$ is
compact and $f$ is elliptic, by applying Lemma \ref{lem3} we obtain 
\begin{equation}
\widetilde{I}\left( j,l,A\right) =\tsum\limits_{i=1}^{\tau }p^{-\left(
M+1\right) n}\Psi \left( p^{j}x\cdot \widetilde{\xi }_{i}\right) \left\vert
f\left( \widetilde{\xi }_{i}\right) \right\vert _{p}^{\beta l}\tint\limits_{%
\mathbb{Z}_{p}^{n}}\Psi \left( p^{j+M+1}x\cdot \eta \right) d\eta ,
\label{I2}
\end{equation}%
where $\tau $, $M=M(f,A)$, \ $\widetilde{\xi }_{i}\in A,$ $i=1,\ldots ,\tau $%
, depend only on $f$ and $A$. Now by using%
\begin{equation*}
\tint\limits_{\mathbb{Z}_{p}}\Psi \left( p^{j+M+1}x_{i}\eta _{i}\right)
d\eta =\left\{ 
\begin{array}{ll}
0, & \text{ if }j<-M-1-ord\left( x_{i}\right) \\ 
&  \\ 
1, & \text{if }j\geq -M-1-ord\left( x_{i}\right) ,%
\end{array}%
\right.
\end{equation*}%
we can rewrite $\widetilde{I}\left( j,l,A\right) $ as 
\begin{equation}
\left\{ 
\begin{array}{ll}
\tsum\limits_{i=1}^{\tau }q^{-\left( M+1\right) n}\Psi \left( p^{j}x\cdot 
\widetilde{\xi }_{i}\right) \left\vert f\left( \widetilde{\xi }_{i}\right)
\right\vert _{p}^{\beta l}, & \text{ if }j\geq -M-1+\max_{i}\left\{
-ord\left( x_{i}\right) \right\} \\ 
&  \\ 
0, & \text{otherwise.}%
\end{array}%
\right.  \label{I3a}
\end{equation}%
We set $\alpha :=\alpha \left( x\right) =-M-1+\max_{i}\left\{ -ord\left(
x_{i}\right) \right\} $, which is a non-negative integer because $\left\Vert
x\right\Vert _{p}=p^{\max_{i}\left\{ -ord\left( x_{i}\right) \right\} }$ can
be taken big enough. With this notation, by combining (\ref{I2a})-(\ref{I3a}%
),

\begin{equation*}
\left\vert I\left( 0,l\right) \right\vert \leq p^{-\left( M+1\right)
n}\left( \tsum\limits_{i=1}^{l}\left\vert f\left( \widetilde{\xi }%
_{i}\right) \right\vert _{p}^{\beta l}\right) \tsum\limits_{j=\alpha
}^{\infty }p^{-j\left( n+\beta dl\right) },
\end{equation*}%
and since $f\left( \xi \right) $ has coefficients in $\mathbb{Z}_{p}$ and $%
\widetilde{\xi }_{i}\in \mathbb{Z}_{p}^{n}$, $i=1,\ldots ,l$,\ 
\begin{equation*}
\left\vert I\left( 0,l\right) \right\vert \leq \left( \frac{l}{1-p^{-\left(
n+\beta d\right) }}\right) \left\Vert x\right\Vert _{p}^{-n}p^{-\alpha \beta
ld}.
\end{equation*}%
By using the previous for $\left\vert I\left( 0,l\right) \right\vert $
estimation in (\ref{I0}),%
\begin{equation*}
\left\vert Z_{0}\left( x,t\right) \right\vert \leq C_{0}\left\Vert
x\right\Vert _{p}^{-n}\left( e^{p^{\left( M+1\right) d\beta }t\left\Vert
x\right\Vert ^{-d\beta }}-1\right) ,
\end{equation*}%
finally, by using the hypothesis $tp^{\left( M+1\right) d\beta }\left\Vert
x\right\Vert _{p}^{-d\beta }\leq 1$, we have 
\begin{equation*}
\left\vert Z_{0}\left( x,t\right) \right\vert \leq Ct\left\Vert x\right\Vert
_{p}^{-d\beta -n}.
\end{equation*}%
$\square $
\end{proof}

\begin{proposition}
\label{prop1} If $p^{\left( M+1\right) d\beta }t\left\Vert x\right\Vert
_{p}^{-d\beta }\leq 1$, then $\left\vert Z\left( x,t\right) \right\vert \leq
Ct\left\Vert x\right\Vert _{p}^{-d\beta -n}$, for $x\in \mathbb{Q}_{p}^{n}$,
and $t>0$.
\end{proposition}

\begin{proof}
By Lemma \ref{lem2}, $\Psi \left( x\cdot \xi \right) e^{-t\left\vert f\left(
\xi \right) \right\vert _{p}^{\beta }}\in L^{1}$ as a function of $\xi $,
for every $x\in \mathbb{Q}_{p}^{n}$, and $t>0$. Then \ by using the
Dominated Convergence Theorem 
\begin{equation*}
Z\left( x,t\right) =\lim_{L\rightarrow \infty }Z_{L}\left( x,t\right)
=\lim_{L\rightarrow \infty }\tint\limits_{\left( p^{-L}\mathbb{Z}_{p}\right)
^{n}}\Psi \left( x\cdot \xi \right) e^{-t\left\vert f\left( \xi \right)
\right\vert _{p}^{\beta }}d\xi .
\end{equation*}%
By a change of variables we have 
\begin{equation*}
Z_{L}\left( x,t\right) =p^{Ln}\tint\limits_{\mathbb{Z}_{p}^{n}}\Psi \left(
p^{-L}x\cdot \xi \right) e^{-p^{L\beta d}t\left\vert f\left( \xi \right)
\right\vert _{p}^{\beta }}d\xi =p^{Ln}Z_{0}\left( p^{-L}x,p^{L\beta
d}t\right) .
\end{equation*}%
Now by applying the Lemma \ref{lema4},

\begin{equation*}
\left\vert Z_{L}\left( x,t\right) \right\vert \leq Cp^{Ln}\left( \frac{%
tp^{L\beta d}}{\left\Vert xp^{-L}\right\Vert _{p}^{n+\beta d}}\right) \leq C%
\frac{t}{\left\Vert x\right\Vert _{p}^{n+\beta d}},
\end{equation*}%
where $C$ is a constant independent of $L$. Therefore 
\begin{equation*}
\left\vert Z\left( x,t\right) \right\vert =\lim_{L\rightarrow \infty
}Z_{L}\left( x,t\right) \leq Ct\left\Vert x\right\Vert _{p}^{-n-\beta d},
\end{equation*}%
if $p^{\left( M+1\right) d\beta }t\left\Vert x\right\Vert _{p}^{-d\beta
}\leq 1$. $\square $
\end{proof}

\begin{theorem}
\label{theo1}For any $x\in \mathbb{Q}_{p}^{n}$, and any $t>0$, 
\begin{equation*}
\left\vert Z\left( x,t\right) \right\vert \leq At\left( \left\Vert
x\right\Vert _{p}+t^{\frac{1}{\beta d}}\right) ^{-d\beta -n},
\end{equation*}%
where $A$ is a positive constant.
\end{theorem}

\begin{proof}
If $t^{\frac{1}{\beta d}}\leq p^{-\left( M+1\right) }\left\Vert x\right\Vert
_{p}$, by applying Proposition \ref{prop1}, we have

\begin{eqnarray*}
\left\vert Z\left( x,t\right) \right\vert &\leq &Ct\left\Vert x\right\Vert
_{p}^{-d\beta -n}\leq Ct\left( \frac{1}{2}\left\Vert x\right\Vert _{p}+\frac{%
1}{2}p^{\left( M+1\right) }t^{\frac{1}{\beta d}}\right) ^{-d\beta -n} \\
&\leq &\frac{2^{d\beta +n}C}{\left( 1+p^{\left( M+1\right) }\right) ^{d\beta
+n}}\frac{t}{\left( \left\Vert x\right\Vert _{p}+t^{\frac{1}{\beta d}%
}\right) ^{d\beta +n}}.
\end{eqnarray*}%
Now if $\left\Vert x\right\Vert _{p}<p^{\left( M+1\right) }t^{\frac{1}{\beta
d}}$, by applying Lemma \ref{lem2}, we have 
\begin{equation*}
C\left( 1+p^{\left( M+1\right) }\right) ^{d\beta +n}t\left( \left\Vert
x\right\Vert _{p}+t^{\frac{1}{\beta d}}\right) ^{-d\beta -n}\geq Ct^{-\frac{n%
}{\beta d}}\geq \left\vert Z\left( x,t\right) \right\vert .
\end{equation*}%
$\square $
\end{proof}

\begin{corollary}
\label{cor1}With the hypothesis of Theorem \ref{theo1}, the following
assertions hold: (1) $Z\left( x,t\right) $ $\in L^{\rho }\left( \mathbb{Q}%
_{p}^{n}\right) $, for $1\leq \rho <\infty $, for any fixed $t>0$; (2) $%
Z\left( x,t\right) $ is a continuous function in $x$, for any fixed $t>0$.
\end{corollary}

\begin{proof}
(1) The first part follows directly from the estimation given in the
previous theorem. (2) The continuity follows from the fact that $Z\left(
x,t\right) $ is the Fourier transform of $e^{-t\left\vert f\left( \xi
\right) \right\vert _{p}^{\beta }}$, $t>0$, that is an integrable function
by Lemma \ref{lem1}.$\square $
\end{proof}

\subsection{Positivity of the Fundamental Solution}

\begin{theorem}
\label{theo1A}For every $x\in \mathbb{Q}_{p}^{n}$, and every $t>0$, $Z\left(
x,t\right) \geq 0$.
\end{theorem}

\begin{proof}
We start by making the following observations about the fiber of $f:\mathbb{Q%
}_{p}^{n}\rightarrow \mathbb{Q}_{p}$ at $\lambda $.

(\textbf{Claim A) } $f^{-1}\left( \lambda \right) $ is a compact subset of $%
\mathbb{Q}_{p}^{n}$.

Since $f$ is continuous $f^{-1}\left( \lambda \right) $ is a closed subset
of $\mathbb{Q}_{p}^{n}$. By applying Lemma \ref{lem1}, 
\begin{equation*}
f^{-1}\left( \lambda \right) \subseteq \left\{ \xi \in \mathbb{Q}%
_{p}^{n}\mid \left\Vert \xi \right\Vert _{p}\leq \left( \frac{\left\vert
\lambda \right\vert }{C_{0}}\right) ^{\frac{1}{d}}\right\} ,
\end{equation*}%
and thus $f^{-1}\left( \lambda \right) $ is a bounded subset of $\mathbb{Q}%
_{p}^{n}$.

(\textbf{Claim B) }The critical set $C_{f}=\left\{ \xi \in \mathbb{Q}%
_{p}^{n}\mid \nabla f\left( \xi \right) =0\right\} $ of the mapping $f$ \ is
reduced to the origin of $\mathbb{Q}_{p}^{n}$.

This claim follows from the Euler identity%
\begin{equation*}
\frac{1}{d}\dsum\limits_{i=1}^{n}\xi _{i}\frac{\partial f\left( \xi \right) 
}{\partial \xi _{i}}=f\left( \xi \right) ,
\end{equation*}%
and the fact that $f$ is an elliptic polynomial.

Now we recall some basic facts about the integration over fibers (see e.g. 
\cite[Sect. 7.6]{Ig2}). We denote by $\omega =d\xi _{1}\wedge \ldots \wedge
d\xi _{n}$ a differential form of degree $n$ on $\mathbb{Q}_{p}^{n}$. Since $%
df$ does not vanish on $\mathbb{Q}_{p}^{n}\setminus \left\{ 0\right\} $,
there exists a $1-$degree differential form $\omega _{0}$, called a
Gelfand-Leray form, such that $\omega =df\wedge \omega _{0}$. Furthermore, $%
\omega _{0}$ does not vanish on $f^{-1}\left( \lambda \right) $, $\lambda
\in \mathbb{Q}_{p}^{n}\setminus \left\{ 0\right\} $. We denote by $\frac{%
d\xi }{df}$ \ the measure induced by the form $\omega _{0}$. \ Now we recall
that for every $\varphi \in \mathcal{S}\left( \mathbb{Q}_{p}^{n}\right) $,%
\begin{equation*}
\dint\limits_{\mathbb{Q}_{p}^{n}}\varphi \left( \xi \right) d\xi
=\dint\limits_{\mathbb{Q}_{p}^{n}\setminus \left\{ 0\right\} }\left(
\dint\limits_{f\left( \xi \right) =\lambda }\varphi \left( \xi \right) \frac{%
d\xi }{df}\right) d\lambda ,
\end{equation*}%
(cf. \cite[Lemma 8.3.2]{Ig2}). Since $\Psi \left( x\cdot \xi \right) \in 
\mathcal{S}\left( \mathbb{Q}_{p}^{n}\right) $, as a function of $\xi $, by
applying integration on fibers to $Z\left( x,t\right) $,%
\begin{equation*}
Z\left( x,t\right) =\dint\limits_{\mathbb{Q}_{p}^{n}\setminus \left\{
0\right\} }e^{-t\left\vert \lambda \right\vert _{p}^{\beta }}\left(
\dint\limits_{f\left( \xi \right) =\lambda }\Psi \left( x\cdot \xi \right) 
\frac{d\xi }{df}\right) d\lambda .
\end{equation*}

Hence in order to prove \ the theorem, it is sufficient to show that%
\begin{equation*}
F\left( \lambda ,x\right) :=\left( \dint\limits_{f\left( \xi \right)
=\lambda }\Psi \left( x\cdot \xi \right) \frac{d\xi }{df}\right) \geq 0\text{%
, for every }x\in \mathbb{Q}_{p}^{n}\setminus \left\{ 0\right\} \text{.}
\end{equation*}

Let $\widetilde{\xi }$ be a fixed point of $f^{-1}\left( \lambda \right) $, $%
\lambda \in \mathbb{Q}_{p}^{n}\setminus \left\{ 0\right\} $. By Claim B we
may assume, after renaming the variables if necessary, that $\frac{\partial f%
}{\partial \xi _{n}}\left( \widetilde{\xi }\right) \neq 0$. \ We set%
\begin{equation*}
y_{j}:=\left\{ 
\begin{array}{lcr}
\xi _{j}, &  & j=1,\ldots ,n-1 \\ 
&  &  \\ 
f\left( \widetilde{\xi }+p^{e}\mathbb{\xi }\right) -f\left( \widetilde{\xi }%
\right) , &  & j=n,%
\end{array}%
\right.
\end{equation*}%
and define the analytic mapping%
\begin{equation*}
\begin{array}{cccc}
\phi : & \mathbb{Z}_{p}^{n} & \rightarrow & \left( p^{l}\mathbb{Z}%
_{p}\right) ^{n} \\ 
&  &  &  \\ 
& \xi & \rightarrow & y.%
\end{array}%
\end{equation*}

By applying the non-Archimedean implicit function theorem (see e.g. \cite[%
Theorem 2.1.1]{Ig2}) there exist $e$, $l\in \mathbb{N}$ such that $y=\phi
\left( \xi \right) $ is a bianalytic mapping \ from $\mathbb{Z}_{p}^{n}$\
onto $\left( p^{l}\mathbb{Z}_{p}\right) ^{n}$. Then $\xi =\phi ^{-1}\left(
y\right) =\left( y_{1},\ldots ,y_{n-1},\dsum\limits_{l=1}^{\infty
}G_{j}\left( y\right) \right) $, where $G_{j}\left( y\right) $ is a form of
degree $j$, and $G_{1}\left( y\right) \neq 0$. By shrinking the
neighborhoods around $\widetilde{\xi }$ and the origin, i.e., by taking $e$
and $l$ big enough, we may assume that the following conditions hold:

\noindent \textbf{(C)} $G_{j}\left( y\right) $ has with coefficients in $%
\mathbb{Z}_{p}$, for every $j$;

\noindent \textbf{(D)}\ the Jacobian $J_{\phi ^{-1}}$of $\phi ^{-1}$
satisfies $\left\vert J_{\phi ^{-1}}\left( y\right) \right\vert
_{p}=\left\vert J_{\phi ^{-1}}\left( 0\right) \right\vert _{p}$, for every $%
y\in \left( p^{l}\mathbb{Z}_{p}\right) ^{n}$;

\noindent \textbf{(E)} $l\geqslant ord\left( x_{n}\right) $, if $x_{n}\neq 0$%
.

Since $f^{-1}\left( \lambda \right) $, $\lambda \in \mathbb{Q}%
_{p}^{n}\setminus \left\{ 0\right\} $, is a compact subset by Claim A, $%
F\left( \lambda ,x\right) $ can expressed as a finite sum of integrals of
the form%
\begin{equation*}
\dint\limits_{\widetilde{\xi }+p^{e}\mathbb{Z}_{p}^{n}\cap f^{-1}\left(
\lambda \right) }\Psi \left( x\cdot \xi \right) \frac{d\xi }{df}.
\end{equation*}%
Now by changing variables $\xi =\phi ^{-1}\left( y\right) $, and using (C),
(D), and (E), we obtain

\begin{equation*}
\dint\limits_{\widetilde{\xi }+p^{e}\mathbb{Z}_{p}^{n}\cap f^{-1}\left(
\lambda \right) }\Psi \left( x\cdot \xi \right) \frac{d\xi }{df}=\left\vert
J_{\phi ^{-1}}\left( 0\right) \right\vert _{p}\dint\limits_{p^{l}\mathbb{Z}%
_{p}^{n-1}}\Psi \left( \dsum\limits_{j=1}^{n-1}x_{j}\xi
_{j}+x_{n}\dsum\limits_{l=1}^{\infty }G_{j}\left( y\right) \right) dy
\end{equation*}%
\begin{equation*}
=\left\vert J_{\phi ^{-1}}\left( 0\right) \right\vert _{p}\dint\limits_{p^{l}%
\mathbb{Z}_{p}^{n-1}}\Psi \left( \dsum\limits_{j=1}^{n-1}x_{j}\xi
_{j}\right) dy=\left( p^{-l\left( n-1\right) }\left\vert J_{\phi
^{-1}}\left( 0\right) \right\vert _{p}\right) \Omega _{-l}\left( x\right)
\geq 0,
\end{equation*}%
where $\Omega _{-l}$ denotes the characteristic function of $p^{-l}\mathbb{Z}%
_{p}^{n-1}$. Therefore $F\left( \lambda ,x\right) \geq 0$.$\square $
\end{proof}

\subsection{Some Additional Results}

We define $\mathcal{B}:=\mathcal{B}\left( \mathbb{Q}_{p}^{n}\right) $ to be
the set of all functions $\varphi :\mathbb{Q}_{p}^{n}\rightarrow \mathbb{C}$
satisfying: (1) $\mathbb{\varphi }$ is continuous; (2) $\left\Vert \varphi
\right\Vert _{L^{\infty }}<\infty $.

\begin{proposition}
\label{prop2}The fundamental solution has the following properties:

\noindent (P1) $\tint\nolimits_{\mathbb{Q}_{p}^{n}}Z\left( x,t\right) $ $%
dx=1 $, for any $t>0$;

\noindent (P2) if $\varphi \in \mathcal{B}$, then $\lim_{\left( x,t\right)
\rightarrow \left( x_{0},0\right) }\tint\nolimits_{\mathbb{Q}%
_{p}^{n}}Z\left( x-\eta ,t\right) \varphi \left( \eta \right) d\eta =\varphi
\left( x_{0}\right) $;

\noindent (P3) $Z\left( x,t+t^{\prime }\right) =\tint\nolimits_{\mathbb{Q}%
_{p}^{n}}Z\left( x-y,t\right) Z\left( y,t^{\prime }\right) dy$, for $t$, $%
t^{\prime }>0$.
\end{proposition}

\begin{proof}
\noindent (P1) It follows from Corollary \ref{cor1} with $\rho =2$ and the
Fourier inversion formula.

\noindent (P2) We set $u\left( x,t\right) =\tint\nolimits_{\mathbb{Q}%
_{p}^{n}}Z\left( x-\eta ,t\right) \varphi \left( \eta \right) d\eta $. Then
we have to show that 
\begin{equation*}
\lim_{\left( x,t\right) \rightarrow \left( x_{0},0\right) }u\left(
x,t\right) =\varphi \left( x_{0}\right) \text{, }
\end{equation*}%
for any fixed $x_{0}\in \mathbb{Q}_{p}^{n}$. Since $\varphi $ is continuous
at $x_{0}$ there exists \ a ball $B:=\left\{ \eta \in \mathbb{Q}_{p}^{n}\mid
\left\Vert \eta -x_{0}\right\Vert _{p}<p^{-e}\right\} $, such that $%
\left\vert \varphi \left( \eta \right) -\varphi \left( x_{0}\right)
\right\vert <\frac{\epsilon }{2}$, for every $\eta \in B$.

Then $\ \left\vert u\left( x,t\right) -\varphi \left( x_{0}\right)
\right\vert \leq \left\vert I_{1}\right\vert +\left\vert I_{2}\right\vert $,
where

\begin{eqnarray*}
\left\vert I_{1}\right\vert &:&=\left\vert \tint\limits_{\left\Vert \eta
-x_{0}\right\Vert <p^{-e}}Z\left( x-\eta ,t\right) \left[ \varphi \left(
\eta \right) -\varphi \left( x_{0}\right) \right] d\eta \right\vert , \\
\left\vert I_{2}\right\vert &:&=\left\vert \tint\limits_{\left\Vert \eta
-x_{0}\right\Vert \geq p^{-e}}Z\left( x-\eta ,t\right) \left[ \varphi \left(
\eta \right) -\varphi \left( x_{0}\right) \right] d\eta \right\vert .
\end{eqnarray*}

By using the continuity of $\varphi $ and (P1), we have

\begin{equation*}
\left\vert I_{1}\right\vert \leq \left\vert \varphi \left( \eta \right)
-\varphi \left( x_{0}\right) \right\vert <\frac{\epsilon }{2},\text{ for }%
\eta \in B.
\end{equation*}

By applying Theorem \ref{theo1} to $\left\vert I_{2}\right\vert $, 
\begin{equation*}
\left\vert I_{2}\right\vert \leq 2Ct\left\Vert \varphi \right\Vert
_{L^{\infty }}\tint\limits_{\left\Vert \eta -x_{0}\right\Vert _{p}\geq
p^{-e}}\left\Vert x-\eta \right\Vert _{p}^{-d\beta -n}d\eta .
\end{equation*}%
Since we are interested in the values of $x$ close to $x_{0}$, we may assume
that $\left\Vert x-x_{0}\right\Vert <p^{-e}$, then by the ultrametric
triangle inequality, 
\begin{equation*}
\left\Vert x-\eta \right\Vert _{p}=\max \left( \left\Vert x-x_{0}\right\Vert
_{p},\left\Vert \eta -x_{0}\right\Vert _{p}\right) =\left\Vert \eta
-x_{0}\right\Vert _{p},
\end{equation*}%
and then 
\begin{equation*}
\left\vert I_{2}\right\vert \leq 2Ct\left\Vert \varphi \right\Vert
_{L^{\infty }}\tint\limits_{\left\Vert \eta \right\Vert _{p}\geq
p^{-e}}\left\Vert \eta \right\Vert _{p}^{-d\beta -n}d\eta \leq
C_{1}t\left\Vert \varphi \right\Vert _{L^{\infty }}\text{, }
\end{equation*}%
for $t>0$, where $C_{1}$\ is a positive constant.

Note that $\left\Vert \varphi \right\Vert _{L^{\infty }}=0$, \ implies $%
\varphi \equiv 0$, since $\varphi $ is a continuous function. In this case
the theorem is valid. For this reason we assume that $\left\Vert \varphi
\right\Vert _{L^{\infty }}>0$. Hence 
\begin{equation*}
\left\vert I_{2}\right\vert <\frac{\epsilon }{2},\text{for }\left( t,\eta
\right) \text{\ satisfying }\left\Vert x-x_{0}\right\Vert <p^{-e}\text{, and 
}0<t<\frac{\varepsilon }{2C_{1}t\left\Vert \varphi \right\Vert _{L^{\infty }}%
}.
\end{equation*}

\noindent (P3) Since $e^{-t\left\vert f\left( \xi \right) \right\vert
_{p}^{\beta }}\in L^{1}$, for every $t>0$, 
\begin{equation*}
\tint\nolimits_{\mathbb{Q}_{p}^{n}}Z\left( x-y,t\right) Z\left( y,t^{\prime
}\right) dy=\mathcal{F}^{-1}\left( e^{-t\left\vert f\left( \xi \right)
\right\vert _{p}^{\beta }}e^{-t^{\prime }\left\vert f\left( \xi \right)
\right\vert _{p}^{\beta }}\right) =Z\left( x,t+t^{\prime }\right) ,
\end{equation*}%
for $t$, $t^{\prime }>0$. $\square $
\end{proof}

\section{The Cauchy Problem for $p-$adic Parabolic Equations}

In this section we study the following Cauchy problem:

\begin{equation}
\left\{ 
\begin{array}{l}
\frac{\partial u\left( x,t\right) }{\partial t}+\left( f\left( D,\beta
\right) u\right) \left( x,t\right) =0,\text{ }t>0\text{,} \\ 
\\ 
u\left( x,0\right) =\varphi \left( x\right) ,%
\end{array}%
\right.  \label{iv1}
\end{equation}%
where $\varphi \in L^{1}\left( \mathbb{Q}_{p}^{n}\right) \cap \mathcal{B}%
\left( \mathbb{Q}_{p}^{n}\right) $.

\begin{lemma}
\label{lema1}If $\varphi \in L^{1}$, then the function 
\begin{equation}
u\left( x,t\right) =\tint\limits_{\mathbb{Q}_{p}^{n}}Z\left( x-\eta
,t\right) \varphi \left( \eta \right) d\eta  \label{u0}
\end{equation}%
satisfies the equation 
\begin{equation*}
\frac{\partial u\left( x,t\right) }{\partial t}+\left( f\left( D,\beta
\right) u\right) \left( x,t\right) =0,t>0.
\end{equation*}%
In addition, $u\left( x,t\right) \in L^{\rho }\left( \mathbb{Q}%
_{p}^{n}\right) $, for $1\leq \rho <\infty $, and for every fixed $t,\beta
>0 $.
\end{lemma}

\begin{proof}
It is clear that one may differentiate in (\ref{u0}) under the integral sign:%
\begin{equation}
\frac{\partial u\left( x,t\right) }{\partial t}=\tint\limits_{\mathbb{Q}%
_{p}^{n}}\varphi \left( \eta \right) \frac{\partial }{\partial t}Z\left(
x-\eta ,t\right) d\eta =\frac{\partial Z\left( x,t\right) }{\partial t}\ast
\varphi \left( x\right) .  \label{u0-1}
\end{equation}%
On the other hand, since $Z\left( x,t\right) \in L^{\rho }$, $1\leq \rho
<\infty $, for any fixed $t>0$ (cf. Corollary \ref{cor1}), and $\varphi \in
L^{1}$, then $u\left( x,t\right) \in L^{\rho }$, $1\leq \rho <\infty $, for
any fixed $t>0$, and its Fourier transform with respect $x$ is $%
e^{-t\left\vert f\left( \xi \right) \right\vert _{p}^{\beta }}\left( 
\mathcal{F}\varphi \right) \left( \xi \right) $. Now by using Lemma \ref%
{lem1} we have $\left\vert f\left( \xi \right) \right\vert _{p}^{\beta
}e^{-t\left\vert f\left( \xi \right) \right\vert _{p}^{\beta }}\in L^{1}\cap
L^{2}$ for any fixed $t>0$. Then $\left( f\left( D,\beta \right)
u_{0}\right) \left( x,t\right) $ is given by 
\begin{eqnarray*}
\left( f\left( D,\beta \right) u\right) \left( x,t\right) &=&\mathcal{F}%
_{\xi \rightarrow x}^{-1}\left( \left\vert f\left( \xi \right) \right\vert
_{p}^{\beta }e^{-t\left\vert f\left( \xi \right) \right\vert _{p}^{\beta
}}\right) \ast \varphi \left( x\right) \\
&=&-\mathcal{F}_{\xi \rightarrow x}^{-1}\left( \frac{\partial }{\partial t}%
e^{-t\left\vert f\left( \xi \right) \right\vert _{p}^{\beta }}\right) \ast
\varphi \left( x\right) ,
\end{eqnarray*}%
for $t>0$, and since one may differentiate in (\ref{fusol}) under the
integral,%
\begin{equation}
\left( f\left( D,\beta \right) u\right) \left( x,t\right) =-\frac{\partial
Z\left( x,t\right) }{\partial t}\ast \varphi \left( x\right) \text{.}
\label{u0-3}
\end{equation}
Now the result follows directly from (\ref{u0-1}) and (\ref{u0-3}). $\square 
$
\end{proof}

\begin{lemma}
\label{lema2}Let $u\left( x,t\right) $ be as in Lemma \ref{lema1}. If $%
\varphi \in \mathcal{B}$, then for all $x\in \mathbb{Q}_{p}^{n}$, and $t>0$,
the inequality%
\begin{equation}
\left\vert u\left( x,t\right) \right\vert \leq \left\Vert \varphi
\right\Vert _{L^{\infty }}  \label{ine}
\end{equation}%
holds.
\end{lemma}

\begin{proof}
By applying Theorem \ref{theo1A}\ and Proposition \ref{prop2} (1), we have 
\begin{equation*}
\left\vert u\left( x,t\right) \right\vert \leq \tint\limits_{\mathbb{Q}%
_{p}^{n}}Z\left( x-\eta ,t\right) \left\vert \varphi \left( \eta \right)
\right\vert d\eta \leq \left\Vert \varphi \right\Vert _{L^{\infty }},
\end{equation*}%
for any $x\in \mathbb{Q}_{p}^{n}$, and $t>0$. $\square $
\end{proof}

\begin{lemma}
\label{lema3}Let $u\left( x,t\right) $ be as in Lemma \ref{lema1}. If $\in
L^{1}\cap \mathcal{B}$, then $u\left( x,t\right) $ is a continuous function
in $x$, for fixed $t$ $\in \left[ 0,T\right] $, $T>0$. Furthermore, $u\left(
x,0\right) =\varphi \left( x\right) $.
\end{lemma}

\begin{proof}
We first note that by Proposition \ref{prop2} (3),%
\begin{equation*}
\lim_{\left( x,t\right) \rightarrow \left( x_{0},0\right) }u\left(
x,t\right) =\varphi \left( x_{0}\right) \text{, }
\end{equation*}%
for any fixed $x_{0}\in \mathbb{Q}_{p}^{n}$. Then it is sufficient to show
that for every fixed $t\in \left( \delta ,T\right] $, where $T\geq \delta >0$
are fixed constants, $u\left( x,t\right) $ is a continuous function in $x$.

We fix $t\in \left( \delta ,T\right] $. Since 
\begin{equation*}
u\left( x,t\right) =\tint\nolimits_{\mathbb{Q}_{p}^{n}}Z\left( \eta
,t\right) \varphi \left( x-\eta \right) d\eta ,
\end{equation*}%
by using a well-known result about the continuity of an integral with
respect to a parameter, it is sufficient to show that $Z\left( \eta
,t\right) \varphi \left( x-\eta \right) $ is continuous in $x,$ for any
fixed $\eta $ and locally integrable in $\eta $ for any fixed $x$, and
further, $\left\vert Z\left( \eta ,t\right) \varphi \left( x-\eta \right)
\right\vert \leq \Omega \left( \eta \right) $ for $\Omega $ in $L^{1}$, for
every $x$. The first condition is a consequence of the fact that $\varphi $
and $Z$ are continuous in $x$ (cf. Corollary \ref{cor1}). The second
condition follows from $\left\vert Z\left( \eta ,t\right) \varphi \left(
x-\eta \right) \right\vert \leq \left\Vert \varphi \right\Vert _{L^{\infty
}}Z\left( \eta ,t\right) \in L^{1}$, for $t\in \left( \delta ,T\right] $,
(cf. Corollary \ref{cor1}). Finally, in order to establish the existence of
the function $\Omega $, we proceed as follows. \ By applying Theorem \ref%
{theo1}, \ we have%
\begin{eqnarray*}
\left\vert Z\left( \eta ,t\right) \varphi \left( x-\eta \right) d\eta
\right\vert &\leq &CT\left\Vert \varphi \right\Vert _{L^{\infty }}\left(
\left\Vert \eta \right\Vert _{p}+t^{\frac{1}{\beta d}}\right) ^{-d\beta -n}
\\
&\leq &CT\left\Vert \varphi \right\Vert _{L^{\infty }}\left( \left\Vert \eta
\right\Vert _{p}\right) ^{-d\beta -n},
\end{eqnarray*}%
if $\left\Vert \eta \right\Vert _{p}\geq 1$. Otherwise,%
\begin{eqnarray*}
\left\vert Z\left( \eta ,t\right) \varphi \left( x-\eta \right) d\eta
\right\vert &\leq &CT\left\Vert \varphi \right\Vert _{L^{\infty }}\left(
\left\Vert \eta \right\Vert _{p}+t^{\frac{1}{\beta d}}\right) ^{-d\beta -n}
\\
&\leq &CT\left\Vert \varphi \right\Vert _{L^{\infty }}\left( \left\Vert \eta
\right\Vert _{p}+\delta ^{\frac{1}{\beta d}}\right) ^{-d\beta -n}.
\end{eqnarray*}%
Therefore, we may define $\Omega \in L^{1}$ as follows:

\begin{equation*}
\Omega \left( \eta \right) =\left\{ 
\begin{array}{llr}
CT\left\Vert \varphi \right\Vert _{L^{\infty }}\left( \left\Vert \eta
\right\Vert _{p}\right) ^{-d\beta -n}, & \text{if } & \left\Vert \eta
\right\Vert _{p}\geq 1 \\ 
&  &  \\ 
CT\left\Vert \varphi \right\Vert _{L^{\infty }}\left( \left\Vert \eta
\right\Vert _{p}+\delta ^{\frac{1}{\beta d}}\right) ^{-d\beta -n}, & \text{if%
} & \left\Vert \eta \right\Vert _{p}<1.%
\end{array}%
\right.
\end{equation*}%
$\square $
\end{proof}

\begin{theorem}
\label{prop3}If $\varphi \in L^{1}\left( \mathbb{Q}_{p}^{n}\right) \cap 
\mathcal{B}\left( \mathbb{Q}_{p}^{n}\right) $, then the Cauchy problem 
\begin{equation*}
\left\{ 
\begin{array}{l}
\frac{\partial u\left( x,t\right) }{\partial t}+\left( f\left( D,\beta
\right) u\right) \left( x,t\right) =0,\text{ }x\in \mathbb{Q}_{p}^{n},\text{ 
}t\in \left( 0,T\right] ,\text{ }T>0, \\ 
\\ 
u\left( x,0\right) =\varphi \left( x\right) ,%
\end{array}%
\right.
\end{equation*}%
has a \ classical solution:%
\begin{equation*}
u\left( x,t\right) =\tint\nolimits_{\mathbb{Q}_{p}^{n}}Z\left( \eta
,t\right) \varphi \left( x-\eta \right) d\eta .
\end{equation*}%
Furthermore, the solution has the following properties:

\noindent (1) $u\left( x,t\right) $ is a continuous function in $x$, for
every fixed $t\in \left[ 0,T\right] $;

\noindent (2) $\sup_{\left( x,t\right) \in \mathbb{Q}_{p}^{n}\times \left[
0,T\right] }\left\vert u\left( x,t\right) \right\vert \leq \left\Vert
\varphi \right\Vert _{L^{\infty }}$;

\noindent (3) $u\left( x,t\right) \in L^{\rho }$, $1\leq \rho <\infty $, for
any fixed $t>0$.
\end{theorem}

\begin{proof}
The result follows from Lemmas \ref{lema1}, \ref{lema2}, \ref{lema3}.
\end{proof}

\section{$p-$adic Markov Processes}

\begin{theorem}
\label{theo3}The fundamental solution $Z\left( x,t\right) $ is a transition
density of a time- and space-homogeneous non-exploding right continuous
strict Markov process without second kind discontinuities.
\end{theorem}

\begin{proof}
By Proposition \ref{prop2} (4) the family of operators

\begin{equation*}
\left( \Theta \left( t\right) f\right) \left( x\right) =\tint\limits_{%
\mathbb{Q}_{p}^{n}}Z\left( x-\eta ,t\right) f\left( \eta \right) d\eta 
\end{equation*}%
has the semigroup property. We know that $Z\left( x,t\right) \geq 0$ and $%
\Theta \left( t\right) $ preserves the function $f\left( x\right) \equiv 1$
(cf. Proposition \ref{prop2} (2)). Thus $\Theta \left( t\right) $ is a
Markov semigroup. The requiring properties of the corresponding Markov
process follow from Proposition \ref{prop2} \ and general theorems of the
theory of Markov processes \cite{D}, see also \cite[Section XVI]{V-V-Z}. $%
\square $
\end{proof}

\end{document}